\documentclass[12pt]{revtex4-1}
\usepackage{epsfig}
\usepackage{graphicx}
\usepackage{caption}
\usepackage{subcaption}
\usepackage{float}
\usepackage{amsmath}
\usepackage{slashed}
\usepackage{color}
\input epsf.tex
\usepackage{amssymb}
\usepackage{amsmath}
\usepackage{amsthm}
\usepackage{amsopn}
\def\comment#1{}

\def\beq{\begin{equation}}
\def\eeq{\end{equation}}
\def\bea{\begin{eqnarray}}
\def\eea{\end{eqnarray}}

\begin{document}

\title{ Dipolar dark matter and CMB B-mode polarization}
\author{S. Mahmoudi$^{1}$\footnote{s.mahmoudi@shirazu.ac.ir}, M. Haghighat$^{1,2}$\footnote{M.Haghighat@shirazu.ac.ir },  S. Modares Vamegh$^{1}$\footnote{ s.modares@shirazu.ac.ir} and R. Mohammadi $^{3,4}$\footnote{rmohammadi@ipm.ir}}
\address{$^1$ Physics Department, College of Sciences, Shiraz University 71454, Shiraz, Iran\\
	$^2$ISC, 71946-94173, Shiraz, Iran\\$^3$Iranian National Science and Technology Museum (INMOST), PO BOX: 11369-14611, Tehran, Iran\\$^4$School of Astronomy, Institute for Research in Fundamental Sciences (IPM), P. O. Box 19395-5531, Tehran, Iran}
\begin{abstract}
We consider dark matter as singlet fermionic particles which carrying magnetic dipole moment to explore its contribution on the polarization of cosmic microwave background
(CMB) photons. We show that Dirac fermionic dark matter has no contribution on the CMB polarization. However, in the case of Majorana dark matter this type of interaction leads to the B-mode polarization in presence of primordial scalar perturbations which is in contrast with standard scenario for the CMB polarization. We numerically calculate the B-mode power spectra and plot $C_l^{BB}$ for different dark matter masses and the $r$-parameter. We show that the dark matter with masses less than 100MeV have valuable contribution on $C_l^{BB}$.  Meanwhile, the dark matters with mass $m_d\leq50MeV$ for $r=0.07$  ( $m_d\leq80MeV$ for $r=0.09$) can be excluded experimentally. Furthermore, our results put a bound on the magnetic dipole moment about $M\leq 10^{-16} e\,\,cm$ in agreement with the other reported constraints.
\end{abstract}
\maketitle
  \section{Introduction}

  The nature of dark matter (DM) and its interactions is one of the most important questions in cosmology and particle physics. In spite of the fact that there are wealth of cosmological evidence for existing dark matter, from galactic clusters and  velocity curves of spining galaxies to gravitational lensing  \cite{DM1,DM2,DM3}, its particle properties has remained elusive.  To explore the nature of dark matter, different experiments have been proposed such as DAMA/LIBRA collaboration at Gran Sasso \cite{direct-detection1}, CoGeNT collaboration at the Soudan Laboratory underground \cite{direct-detection2} and CDMS collaboration \cite{direct-detection3} which are introduced to detect dark matter directly. In these experiments the scattering of dark matter from nucleons can be described by multipole interactions.  In fact, a dark matter has zero electric charge and therefore in the simplest extension of standard model it can be coupled to photon through an intrinsic electric and or magnetic dipole moments which is well-known as dipolar dark matter (DDM) model\cite{DDM.EM,DDM3,DDM4,DDM5,DDM6}. However, the DDM model can successfully explain some claims of DAMA/LIBRA  and COGENT collaborations\cite{DDM1,DDM2}.

 The CMB photons is expected to be linearly polarized due to the anisotropic Compton scattering around the epoch of recombination. Meanwhile, according to the standard scenario of cosmology there is no physical mechanism to generate a circular polarized radiation at the last scattering surface. However, studies conducted in recent years show that the interaction between photon and matter can convert or generate the polarization states of photon in different situations. For instance, the linear polarization of the CMB photons can be converted to the circular polarization in the presence of background fields or due to the effects of particle scattering which has been widely discussed in the literature \cite{circ1,circ2,circ3,circ4,circ5,circ6,circ7,circ8}.    In this paper, we consider the DDM model with a singlet spin $\frac{1}{2}$ fermion as the  dark matter to examine the effects of magnetic dipole moments on the CMB photon polarization.

   Generally, the CMB polarization pattern has two geometrical components, E-mode and B-mode. These modes based on the Stokes parameters U and Q  can form an independent local coordinate system \cite{stoke1,stoke2,stoke3,stoke4,stoke5}. According to the standard model of cosmology, while E-mode polarization of Compton scattering at the last scattering surface can be produced due to the scalar and tensor perturbations, its B-mode polarization can only be produced by the tensor perturbations. Nevertheless, it has been shown that it is also possible to produce the B-mode polarization in the presence of scalar perturbations. Since the detection of B-mode polarization can provide a unique tool to investigate the CMB perturbations, it is important to identify all potential sources of the B-mode polarization.  As the new sources, for instance, in \cite{B polarization of the CMB from Farady rotation} the effects of the Faraday rotation due to the uniform magnetic field on the CMB is investigated and it is shown that a nonvanishing B-mode can be produced through Farady rotation.  In \cite{neutrino}, the authors have discussed that photon-neutrino interaction in the presence of scalar perturbations could be considered as one of the sources of the CMB B-mode polarization. It is also shown that the Compton scattering in the non-commutative space time can generate the B-mode polarization of the CMB \cite{non commutative}.

    However, the parameter which characterizes the amplitude of metric tensor perturbation is  $r=P_{T}/P_{S}$ where  $P_{T}=A_{T}(k/k_{\circ})^ {n_{T}-1}$ and $P_{S}=A_{S}(k/k_{\circ})^ {n_{S}-1}$ are, respectively, the power spectra of tensor and scalar metric perturbations and $n_{T,S}$ and $A_{T,S}$ are their spectral indices and amplitudes. The $r$ parameter is usually calculated by comparing the B-mode and E-mode power spectra. Recent measurements of BICEP2 + Keck Array + Planck (BKP) report an upper bound $r_{0.002} < 0.09$ \cite{bicep}.

  In this work, we will show that magnetic like component of the CMB polarization  (B-mode polarization) can be produced by the photon-DM interaction in the presence of scalar perturbations. The paper is organized as follows: we introduce the effective Lagrangian for the interaction of dipolar dark matter with photons in section 2. Then we give a brief introduction to the stokes parameters and drive the time evolution of these parameters in terms of the photon-DM scattering in section 3.  The power spectrum is evaluated numerically in section 4.  We compare our results with the experimental data and give some discussion in section 5.

\section{Dipolar Dark Matter Model}

A particle as a candidate for the dark matter is generally known as a stable or relatively stable particle that does not interact electromagnetically. However in recent years, there are some interest in the study of electromagnetic interactions of DM. Such a particle has not probably the electric charge otherwise it has a significant interaction with the photons and could be easily detected. But this particle can weakly couple with the electromagnetic field through loop corrections. In fact, the most general form for the electromagnetic current between fermions consistent with the Lorentz covariance and the Ward identity can be written as follows\cite{ddm}:
\begin{eqnarray}
J_{\mu}^{em}&=&\bar{\psi}(p^{'})\bigg[\gamma_{\mu}F_{1}(q^2)-\gamma_{2}\gamma_{5}\big(g_{\mu}^{\lambda}q^2-q^{\lambda}q_{\mu}\big)G_{1}(q^2)+\nonumber\\
&&\sigma_{\mu \nu}q^{\nu}\big[F_{2}(q^2)+\gamma_{5}G_{2}(q^2)\big]\bigg]\psi(p),
\end{eqnarray}
where $F_1$, $G_1$, $F_2$ and $G_2$ are the electric, anapole, magnetic and electric dipole form factors, respectively.  The current $J_{\mu}^{em}$ can be coupled with photons where its dipolar part is given by
 \begin{equation}\label{DDM2}
 \mathcal{L}_{\rm DDM}=-\frac{i}{2}\bar{\psi}\sigma_{\rm \mu\nu}(M+\gamma^5 D)\psi F^{\mu\nu},
 \end{equation}
  where $F^{\mu\nu}$ is the electromagnetic field, $M$ and $D$ are magnetic and electric dipole moment, respectively and $\sigma^{\mu\nu}=\frac{i}{2}[\gamma^\mu,\gamma^\nu]$.

It should be noticed that the permanent dipole moment can be defined just for Dirac particle and Majorana particle can not have permanent dipole moment. However, Majorana fermions have only nonzero transition moments between different mass eigenstates. Their interactions with photons is described by \cite{DDM3}
    \begin{equation}\label{DDM21}
    \mathcal{L}_{\rm DDM}=-\frac{i}{2}\bar{\psi}_{2}\sigma_{\rm \mu\nu}(M_{12}+\gamma^5 D_{12})\psi_{1} F^{\mu\nu} + H.c.
    \end{equation}
    where $M_{12}$ is a transition magnetic moment and $D_{12}$ is a transition electric moment.
  The Lagrangian (\ref{DDM2}) and  (\ref{DDM21}) form the basis of the DDM model \cite{DDM.EM}. Therefore, the fermionic DM-particle can interact  with photons via electric and magnetic dipole moments.

\section{TIME EVOLUTION OF STOKES PARAMETERS DUE TO DDM-PHOTON SCATTERING}

CMB temperature anisotropy via Compton scattering in the epoch of recombination can cause the polarization of photons. One of the usual methods to characterize  the polarization state of radiation field is throughout the Stokes parameters I, Q, U and V. To introduce these parameters in the context of quantum mechanics, one can consider a photon ensemble. The polarization density matrix of photons is defined as

\begin{equation}\label{stoke6}
\rho = \frac{1}{2}
\Bigg[
\begin{array}{rr}
I+Q \,\,& U-iV \\\\
U+iV\, & I-Q
\end{array}
\Bigg],
\end{equation}
where I is the total intensity of radiation, U, Q and V describe the polarization of photon and for unpolarized photon $Q=U=V=0$. The circularly polarized radiation is defined by none-zero value for V and the linear polarization is described by the Stokes parameters Q and U. The parameters I and V are independent of reference frame whereas Q and U are frame dependent.  Therefore, in the context of cosmology by introducing a set of linear combination of Q and U, one can find reference frame independent parameters that are known as E and B modes.

Meanwhile, time evolution of the Stokes parameters can be examined through the Boltzmann equation. This equation provides a systematic way to account for different couplings in a system and is generally expressed as follows
\begin{equation}\label{Boltz}
	\frac{d f}{dt}= C[f],
\end{equation}
where $C[f]$ in the right-hand side of (\ref{Boltz}) contains all possible collision terms while the left-hand side is known as the Liouville term and
 involves the effects of gravitational perturbations about
the homogeneous cosmology. In the case of photon, the
distribution function $f$ is the density matrix $\rho_{ij}$ as is given in (\ref{stoke6}).
Thus the density operator corresponding to the density matrix $\rho_{ij}$ can be given as
\begin{equation}\label{stoke7}
\hat{\rho}=\frac{1}{tr(\hat{\rho})}\int\frac{d^3p}{(2\pi)^3}\rho_{ij}(p)a_i^\dagger(p)a_j(p),
\end{equation}
and the number operator $D_{ij}^0(k)=a_i^\dagger(k)a_j(k)$, has an expectation value as fallows
\begin{equation}\label{stoke8}
\langle D_{ij}^0(k)\rangle\equiv tr[\rho\hat{ D}_{ij}^0(k)]=(2\pi)^3 \delta^3(0)(2k^0)\rho_{ij}(k).
\end{equation}
However, to examine the time evolution of the photons  polarization in the CMB, we need the time evolution of the density matrix. To this end, we substitute (\ref{stoke8}) in
\begin{equation}\label{stoke9}
\frac{d}{dt}D_{ij}^0(k)=i[H,D_{ij}^0(k)],
\end{equation}
where $H$ is the full Hamiltonian, to find the time evolution of $\rho_{ij}$  as
\begin{equation}\label{stoke10}
(2\pi)^3 \delta^3(0)(2k^0)\frac{d}{dt}\rho_{ij}(k)=i\langle[H_I^0(t),D_{ij}^0(k)]\rangle-\frac{1}{2}\int dt \langle[H_I^0(t),[H_I^0(0),D_{ij}^0(k)]]\rangle.
\end{equation}
In (\ref{stoke10}) $H_I^0 $ is the interacting Hamiltonian at the lowest order\cite{stoke1}. The first and the second term on the right-handed side of (\ref{stoke10}) are called forward scattering and higher order collision term, respectively.
\subsection{Dirac Fermionic Dark Matter}
Firstly,we consider a Dirac fermionic dark matter which interacts with photon via its magnetic dipole moment with the following Lagrangian
\begin{equation}\label{DDM3}
\mathcal{L}_{\rm DDM}=-\frac{i}{2}M\bar{\psi}\sigma_{\mu\nu}\psi F^{\mu\nu}.
\end{equation}
The Feynman diagram corresponding to DDM-photon scattering at the lowest order is very similar to the Compton scattering as is shown in Fig.1. Therefore, the interacting Hamiltonian at the lowest order can be obtained as follows
\begin{eqnarray}\label{DDM4}
H_{I}(t)&=&-M^2 \int d^4 x'\int d^3 x \bar{\psi}^- (x)\sigma^{\mu\nu} S_{F}(x-x')\sigma^{\alpha\beta} (\partial_{\mu}A_{\nu}^- (x)\partial_{\alpha}A_{\beta}^+ (x') \nonumber\\
 &+&\partial_{\alpha}A_{\beta}^- (x')\partial_{\mu}A_{\nu}^+ (x))\psi^+ (x'),
\end{eqnarray}
\begin{figure}
  \includegraphics[width=4in]{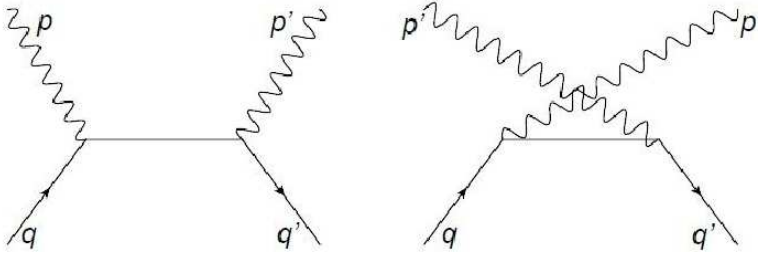}\\
  \caption{The typical diagrams for photon-dark matter scattering}\label{E}
\end{figure}

 with the Fourier transformations of the fields and propagator  as follows
 \begin{equation}\label{DDM5}
 A_{\mu}(x)=\int\frac{d^3 \bf{k}}{(2\pi)^3 2k^0}[a_{s}(k)\epsilon_{s\mu}(k)e^{-ik.x} + a_{s}^\dagger (k)\epsilon_{s\mu}^* (k)e^{ik.x}],
 \end{equation}
 \begin{equation}\label{DDM6}
 \psi_{f}(x)=\int\frac{d^3\bf{q}}{(2\pi)^3}\frac{m_{f}}{q^0}\sum_{r}^{}[b_{r}(q)\mathcal{U}_{r}(q)e^{-iq.x} + d_{r}^\dagger (q)\mathcal{V}_{r}(q)e^{iq.x}],
 \end{equation}
 and
 \begin{equation}\label{DDM7}
 S_{F}(x)=\int \frac{d^4 k}{(2\pi)^4}\frac{\slashed{k} + m}{k^2 - m^2 + i\theta} e^{-ik.x}.
 \end{equation}
 where $\epsilon_{s\mu}(k)$'s are the photon polarization 4-vectors with $s=1,2$ for two physical transverse polarization of a free photon and $a_{s}(k)$($a_{s}^\dagger (k)$) is the annihilation (creation) operator which satisfies the canonical commutation relation as
 \begin{equation}\label{DDM8}
 	[a_{s}(k),a_{s'}^\dagger (k')]=(2\pi)^3 2k^0 \delta_{ss'}\delta^{(3)}(k-k'),
 \end{equation}
 In (\ref{DDM6}) $\mathcal{U}_{r}$ and $\mathcal{V}_{r}$ are the Dirac spinors, $b_{r}$($d_{r}$) and $b_{r}^\dagger$ $(d_{r}^\dagger)$ are, respectively, the annihilation and creation operators for fermion (antifermion) satisfying
 \begin{equation}\label{DDM9}
 	\{b_{r}(q),b_{r'}^\dagger (q')\} =\{d_{r}(q),d_{r'}^\dagger (q')\}=(2\pi)^3 \frac{q^0}{m} \delta_{rr'} \delta^{(3)}(q-q').
 \end{equation}
Therefore, the interaction Hamiltonian (\ref{DDM4}) by using (\ref{DDM5}), (\ref{DDM6}) and (\ref{DDM7}) can be cast into
 \begin{eqnarray}\label{DDM10}
 	H_{I}(t)=\int {d\bf{q}} {d\bf{q'}} {d\bf{p}} {d\bf{p'}} &&(2\pi)^3 \delta^{(3)} ({\bf{q}}+{\bf{p}}-{\bf{q'}}-{\bf{p'}}) e^{i(q^0 +p^0 -q'^0 - p'^0)}\nonumber\\
 	&& [b_{r'}^\dagger(q') a_{s'}^\dagger(p') \mathcal{M} a_{s}(p) b_{r}(q)],
 \end{eqnarray}
 where ${d \bf{q}} \equiv \frac {d^3 \bf{q}}{(2\pi)^3}\frac{m}{q^0}$ and ${d \bf{p}} \equiv \frac {d^3 \bf{p}}{(2\pi)^3}\frac{1}{2 p^0}$ and
  \begin{equation}
 \mathcal{M}\equiv \mathcal{M}_{1} +\mathcal{M}_{2},
 \end{equation}
 with
 \begin{equation}\label{DDM11}
 \mathcal{M}_{1}=M^2 \frac{\bar{\mathcal{U}}_{r'}(q') \slashed{\epsilon}_{s'}(p') \slashed{p'} (\slashed{q}+\slashed{p}+m) \slashed{\epsilon}_{s}(p) \slashed{p} \mathcal{U}_{r}(q)}{2q.p},
 \end{equation}
 and
 \begin{equation}\label{DDM12}
 \mathcal{M}_{2}=-M^2 \frac{\bar{\mathcal{U}}_{r'}(q') \slashed{\epsilon}_{s}(p) \slashed{p} (\slashed{q}-\slashed{p'}+m) \slashed{\epsilon}_{s'}(p') \slashed{p'} \mathcal{U}_{r}(q)}{2q.p'}.
 \end{equation}
  Now, we are ready to evaluate the forward scattering term, the first term on the right hand side of (\ref{stoke10}). For this purpose, one needs the expectation value of operators such as \cite{stoke1}
 \begin{equation}\label{DDM13}
 	\langle a_{1}a_{2}...b_{1}b_{2}...\rangle=\langle a_{1}a_{2}...\rangle \langle b_{1}b_{2}...\rangle,
 \end{equation}
 \begin{equation}\label{DDM14}
 	\langle a_{s'}^\dagger(p') a_{s}(p)\rangle = 2p^0 (2\pi)^3 \delta^{(3)}({\bf{p}}-{\bf{p'}})\rho_{ss'}({\bf{x}},{\bf{p}}),
 \end{equation}
 and
\begin{equation}\label{DDM15}
	\langle b_{r'}^\dagger(q') b_{r}(q)\rangle = (2\pi)^3 \delta^{(3)}({\bf{q}}-{\bf{q'}})\delta_{ss'}n_{d}({\bf{x}},{\bf{q}}),
\end{equation}
to evaluate $ \langle [H_{I}^0(0), D_{ij}({\bf{k}})]\rangle$ as follows
\begin{equation}\label{DDM16}
i\langle [H_{I}^0(0), D_{ij}({\bf{k}})]\rangle = i\int d{\bf{q}}n_{d}({\bf{x}},{\bf{q}})(\delta_{is}\rho_{s'j}(k)-\delta_{js'}\rho_{is}(k))(2\pi)^3 \delta^{(3)}(0)\mathcal{M},
\end{equation}	
where
\begin{equation}\label{DDM17}
\mathcal{M}= M^2\bar{\mathcal{U}}_{r}(q)(\slashed{\epsilon}_{s'}(k)\slashed{\epsilon}_{s}(k)-\slashed{\epsilon}_{s}(k)\slashed{\epsilon}_{s'}(k))\slashed{k} \mathcal{U}_{r}(q),
\end{equation}
or
\begin{equation}\label{DDM18}
\mathcal{M}=2 M^2 \epsilon _{s'}^\alpha \epsilon_{s}^\beta k^\sigma \epsilon_{\alpha\beta\sigma\lambda}\bar{\mathcal{U}}_{r}(q)\gamma^\lambda\gamma^5 \mathcal{U}_{r}(q).
\end{equation}
Meanwhile, in the nonrelativistic limit one has \cite{29}
\begin{equation}\label{DDM19}
\bar{\mathcal{U}}_{r}(q)\gamma_{0}\gamma_{5}\mathcal{U}_{s}(q)\simeq-\xi_{r}^\dagger\frac{\vec{q}.\vec{\sigma}}{m_{f}}\xi_{s},
\end{equation}
and
\begin{equation}\label{DDM20}
\bar{\mathcal{U}}_{r}(q)\gamma_{i}\gamma_{5}\mathcal{U}_{s}(q)\simeq-\xi_{r}^\dagger\sigma_{i}\xi_{s},
\end{equation}
where $\xi$ is the two component spinor normalized to unity.
Therefore, after some manipulations the amplitude can be rewritten as
\begin{equation}\label{DDM22}
\mathcal{M}\simeq -2 M^2\{\vec{k}.(\vec{\epsilon}_{s'}\times\vec{\epsilon}_{s})\xi_{r}^\dagger\frac{\vec{q}.\vec{\sigma}}{m_{f}}\xi_{r} +k^0 (\vec{\epsilon}_{s'}\times\vec{\epsilon}_{s}).\xi_{r}^\dagger\vec{\sigma}\xi_{r}\}.
\end{equation}
Unfortunately, (\ref{DDM22}) for the Dirac fermions  with both helicity degree of freedom (left and right handed helicity), leads to a vanishing average on the fermion helicity $r$ as
\begin{equation}\label{r-av}
   \frac{1}{2}\sum_{r}\Big(-2 M^2\{\vec{k}.(\vec{\epsilon}_{s'}\times\vec{\epsilon}_{s})\xi_{r}^\dagger\frac{\vec{q}.\vec{\sigma}}{m_{f}}\xi_{r} +k^0 (\vec{\epsilon}_{s'}\times\vec{\epsilon}_{s}).\xi_{r}^\dagger\vec{\sigma}\xi_{r}\}\Big)=0.
\end{equation}
In fact, the photon-dark matter forward scattering for a Dirac fermions with both handedness has not any contribution on the CMB polarization.
\subsection{Majorana Dark Matter}
Now, we investigate the effect of dipolar dark matter-photon interaction  on the CMB polarization through Majorana magnetic moment. In fact, we are going to consider right handed strile neutrinos which interact with photons through Majorana magnetic moment based on the following Lagrangian \cite{majorana}
  \begin{equation}\label{majorana}
  	\mathcal{L}=-\frac{i}{2} \,\,M_{12}\,\,\bar{\chi^{c}}_{2}\sigma_{\rm \mu\nu} P_{R}\chi_{1}\,\,F^{\mu\nu}\,\,+H.c. ,
  \end{equation}
  where $P_{R}=\frac{1}{2}(1+\gamma^{5})$ , $\bar{\chi^{c}}=(-i\gamma_{2}\chi^{\star})^{\dagger}\gamma_{0}$. Using this Lagrangian, the interacting Hamiltonian at the lowest order can be obtained as follows
     \begin{eqnarray}\label{majorana2}
   H_{I}(t)&=&\frac{-i}{2}\,\,M_{12}^{2} \int d^4 x'\int d^3 x\,\, \bar{\chi}^{-}_{1}(x)\sigma^{\mu\nu} P_{R}\,\,S_{F}^{c}(x-x')\,\,P_{L}\sigma^{\alpha\beta} (\partial_{\mu}A_{\nu}^- (x)\partial_{\alpha}A_{\beta}^+ (x') \nonumber\\
   &+&\partial_{\alpha}A_{\beta}^- (x')\partial_{\mu}A_{\nu}^+ (x))\chi^{+}_{1} (x'),
   \end{eqnarray}
   It is important to mention that in this article, we have assumed that one of the two Majorana particle can decay into another one and therefore we have considered the stable one as a Majorana dark matter.
  Using the Fourier transformation of the fields and propagators
  \begin{equation}\label{majorana3}
  A_{\mu}(x)=\int\frac{d^3 \bf{k}}{(2\pi)^3 2k^0}[a_{s}(k)\epsilon_{s\mu}(k)e^{-ik.x} + a_{s}^\dagger (k)\epsilon_{s\mu}^* (k)e^{ik.x}],
  \end{equation}
  \begin{equation}\label{majorana4}
  \chi(x)=\int\frac{d^3\bf{q}}{(2\pi)^3}\frac{m_{f}}{q^0}\sum_{r}^{}[b_{r}(q)\mathcal{U}_{r}(q)e^{-iq.x} + d_{r}^\dagger (q)\mathcal{V}_{r}(q)e^{iq.x}],
  \end{equation}
   \begin{equation}\label{majorana7}
  S_{F}^{c}(x)=\int \frac{d^4 k}{(2\pi)^4}\frac{\slashed{k} - m}{k^2 - m^2 + i\theta} e^{-ik.x},
  \end{equation}
  we arrive at the following relation for the Feynman amplitude $\mathcal{M}^{'}$:
    \begin{eqnarray}\label{majorana9}
   \mathcal{M}^{'}&=& \frac{-1}{2}M_{12}^2\,\, \Big(\frac{\bar{\mathcal{U}}_{r'}(q') \slashed{\epsilon}_{s'}(p') \slashed{p'}\,\,P_{L} (\slashed{q}+\slashed{p}-m_{2}) \slashed{\epsilon}_{s}(p) \slashed{p}\,\,P_{R}\,\, \mathcal{U}_{r}(q)}{(p+q)^{2}-m_{2}^{2}}+\nonumber\\
   && \frac{\bar{\mathcal{U}}_{r'}(q') \slashed{\epsilon}_{s}(p) \slashed{p} \,\,P_{L}(\slashed{q'}-\slashed{p}-m_{2}) \slashed{\epsilon}_{s'}(p') \slashed{p'}\,\,P_{R}\,\, \mathcal{U}_{r}(q)}{(q'-p)^{2}-m_{2}^{2}}\Big)\nonumber\\\\
   &&=\frac{-1}{2} M_{12}^2\,\, \Big(\frac{\bar{\mathcal{U}}_{r'}(q') \slashed{\epsilon}_{s'}(p') \slashed{p'}\,\, (\slashed{q}+\slashed{p}) \slashed{\epsilon}_{s}(p) \slashed{p}\,\,P_{R}\,\, \mathcal{U}_{r}(q)}{2q.p-(m_{2}^{2}-m_{1}^{2})}+\nonumber\\
   && \frac{\bar{\mathcal{U}}_{r'}(q') \slashed{\epsilon}_{s}(p) \slashed{p} \,\,(\slashed{q'}-\slashed{p}) \slashed{\epsilon}_{s'}(p') \slashed{p'}\,\,P_{R}\,\, \mathcal{U}_{r}(q)}{-2p.q'- (m_{2}^{2}-m_{1}^{2})}\Big).
  \end{eqnarray}
  To evaluate the forward scattering term, one needs to calculate $ \langle [H_{I}^0(0), D_{ij}({\bf{k}})]\rangle$ as follows
  \begin{equation}\label{DDM25}
  i\langle [H_{I}^0(0), D_{ij}({\bf{k}})]\rangle = i\int d{\bf{q}}n_{d}({\bf{x}},{\bf{q}})(\delta_{is}\rho_{s'j}(k)-\delta_{js'}\rho_{is}(k))(2\pi)^3 \delta^{(3)}(0)\mathcal{M}^{'},
  \end{equation}	
  where
  \begin{equation}
  	\mathcal{M}^{'}=\mathcal{M}^{'}\,\,\vert_{q'=q,\,\,p'=p=k},
  \end{equation}
   hence
  \begin{eqnarray}\label{majorana11}
  	\mathcal{M}^{'}&=&\frac{-1}{2}\Big( M_{12}^{2}\,\,\frac{(2k.q)^{2}}{(2k.q)^{2}-(m_{2}^{2}-m_{1}^{2})^{2}}\bar{\mathcal{U}}_{r}(q)(\slashed{\epsilon}_{s'}(k)\slashed{\epsilon}_{s}(k)-\slashed{\epsilon}_{s}(k)\slashed{\epsilon}_{s'}(k))\slashed{k}\,\,P_{R}\,\,\mathcal{U}_{r}(q)\Big),
  \end{eqnarray}
  or
   \begin{eqnarray}\label{majorana12}
   	\mathcal{M}^{'}&=& -M_{12}^{2}\Big(\,\,\frac{(2k.q)^{2}}{(2k.q)^{2}-(m_{2}^{2}-m_{1}^{2})^{2}}\,\,\epsilon _{s'}^\alpha \epsilon_{s}^\beta k^\sigma \epsilon_{\alpha\beta\sigma\lambda}\,\,\bar{\mathcal{U}}_{r}(q)\gamma^\lambda\mathcal{U}_{r}(q)\Big).
   \end{eqnarray}
  By using identity
  \begin{equation}\label{majorana13}
  \bar{\mathcal{U}}_{r}(q)\gamma^{\lambda}\,\,\mathcal{U}_{s}(q)=2\,\,\frac{q^{\lambda}}{m}\,\delta_{rs},
  \end{equation}
      (\ref{majorana12}) cast into
    \begin{eqnarray}\label{majorana15}
    \mathcal{M}^{'}&=& -M_{12}^{2}\,\,\frac{(2k\cdot q)^{2}}{(2k.q)^{2}-(m_{2}^{2}-m_{1}^{2})^{2}}\,\,\Big(\frac{q^{0}}{m_{1}}\vec{k}.(\vec{\epsilon}_{s'}\times\vec{\epsilon}_{s})-k^{0}\,\,v_{b}(\vec{\epsilon}_{s'}\times\vec{\epsilon}_{s})\cdot\vec{\hat{v}}_{b}\Big).
    \end{eqnarray}
   \subsubsection{ In the case $\delta=m_2-m_1<<k^0$}
   Here, we consider the case that  $\delta=m_{2}-m_{1}\ll k^{0}$ and therefore (\ref{majorana15}) can be estimated as follows
\begin{equation}\label{DDM23}
	\mathcal{M}\simeq - M^2\,\,\vec{k}.(\vec{\epsilon}_{s'}\times\vec{\epsilon}_{s}),
\end{equation}
where $M_{12}=M$ and $v_b=|\vec{q}|/m_f$ is the bulk velocity of dark matter  [for example see \cite{Gramann:1997rh}].  Although the second term in (\ref{DDM23}) is similar to the first one, a straightforward calculation leads to a negligible value for this term. In fact, the order of the second term is smaller than the first one due to the presence of $v_{b}$ and therefore we ignore the secend term of (\ref{DDM23}).  Now by substituting  (\ref{DDM23}) in (\ref{DDM25}) and using (\ref{stoke10}), the time evolution of density matrix element can be written as
\begin{equation}\label{DDM24}
	\frac{d\rho_{ij}}{dt}=- i M^2\int d{\bf{q}}n_{d}({\bf{x}},{\bf{q}})(\delta_{is}\rho_{s'j}(k)-\delta_{js'}\rho_{is}(k))(\vec{\epsilon}_{s'}\times\vec{\epsilon}_{s})\cdot \hat{k},
\end{equation}
where $\hat{k}=\vec{k}/k^0$. Consequently, the stokes parameters evolve as
\begin{equation}\label{DD25}
	\frac{d I}{dt}=C_{e\gamma}^I,
\end{equation}
\begin{equation}\label{DDM26}
	\frac{d}{dt}(Q\pm iU)=C_{e\gamma}^\pm \mp i\dot{\tau}_{d}(Q\pm iU),
\end{equation}
\begin{equation}\label{DDM27}
	\frac{d V}{dt}=C_{e\gamma}^V,
\end{equation}
where $C_{e\gamma}^I$,$C_{e\gamma}^V$ and $C_{e\gamma}^\pm$ show the contribution of Thomson scattering \cite{stoke1} and  $\dot{\tau}_{d}$ is defined as follow
\begin{equation}\label{DDM28}
	\dot{\tau}_{d}=\frac{3}{8\pi}(\frac{m_{e}}{\alpha})^2 \sigma_{T}\, M^2\, \,n_{d} ,
\end{equation}
where $\sigma_{T}$ is the Thomson cross section and the dark matter number density $n_{d}$ is
\begin{equation}\label{DDM29}
n_{d}({\bf{x}})=\int \frac{d^3 q}{(2\pi)^3}\frac{m_{f}}{q_{0}}n_{d}({\bf{x}},{\bf{q}}).	
\end{equation}
It should be noted that the second term in the right hand side of (\ref{DDM26}) which comes from the photon-DDM forward scattering affects the time evolution of the stokes parameters $Q$ and $U$.

 To evaluate this term one needs the relation between magnetic dipole moment $M$ and the dark matter-photon scattering cross section $\langle \sigma v\rangle$ \cite{DDM.EM,DDM3}
\begin{equation}\label{DDM30}
\langle \sigma v\rangle\approx\frac{1}{2\pi} M^4 m_{d}^2 ,
\end{equation}
 which cast (\ref{DDM28}) into
 \begin{equation}\label{DDM31}
 \dot{\tau}_{d}=\,\frac{3}{8\pi}\,\,(\frac{m_{e}}{m_{d}})^2 \,\,\frac{\sigma_{T}}{\alpha^2} \sqrt{2\pi\langle\sigma v\rangle}\,\, \rho_{d} ,
 \end{equation}
 where $\rho_{d}$ is the dark matter mass density.
Since the number density of electron is equal to the number density of proton and it is approximately equal to the baryonic matter number density
\begin{equation}
n_{e} = n_{p} \approx n_{B.M}
\end{equation}
 then the ratio of $\dot{\tau}_{d}$ with respect to the  $\dot{\tau}_{e}$ corresponding to the Thomson cross section $\sigma_{T}$ can be found as
	\begin{equation}\label{DDMM}
		\frac{\dot{\tau}_{d}}{\dot{\tau}_{e}}=\,\frac{3}{8\pi}\,\,(\frac{m_{e}}{m_{d}})^2 \,\,\frac{\Omega_{d}}{\Omega_{B.M}} \,\, \frac{m_{p}}{\alpha^2}\,\, \sqrt{2\pi\langle\sigma v\rangle},
	\end{equation}
	where $\Omega_{d}$ and $\Omega_{B.M}$ are the dark matter density parameter and the baryonic matter density parameter, respectively.
	 However, value of the ratio given in (\ref{DDMM}) can be estimated as
	\begin{equation}
		\frac{\dot{\tau}_{d}}{\dot{\tau}_{e}}\simeq 5.2\times10^{-11}\,\, \,\,    (\frac{m_d}{10 GeV})^{-2}\,\,\,\,\,
		     (\frac{\langle\sigma v\rangle}{(10^{-30})\frac{cm^3}{s}})(\frac{\Omega_{d}}{0.26})(\frac{\Omega_{B.M}}{0.04})(\frac{m_p}{1 GeV}),
	\end{equation}
	where for dark matter with masses $10 GeV$ \textemdash $10 MeV$ varies as $5.2\times10^{-11}$ \textemdash $5.2\times10^{-5}$.
\subsubsection{ In the case $\delta=m_2-m_1>>k^0$}
	In this case, (\ref{majorana15}) can be estimated as follows
	 \begin{eqnarray}\label{majorana16}
	 \mathcal{M}^{'}\simeq M_{12}^{2}\,\,\frac{(2k\cdot q)^{2}}{(m_{2}^{2}-m_{1}^{2})^{2}}\,\,\vec{k}\cdot(\vec{\epsilon}_{s'}\times\vec{\epsilon}_{s}),
	 \end{eqnarray}
	In the nonrelativistic limit and if we assume $m_{1}\approx m_{2}$, we will have
	\begin{eqnarray}\label{majorana17}
	\mathcal{M}^{'}\simeq M_{12}^{2}\,\,(\frac{k^{0}}{\delta})^{2}\,\,\vec{k}\cdot(\vec{\epsilon}_{s'}\times\vec{\epsilon}_{s}),
	\end{eqnarray}
	For the cases in which $k^{0}<<\delta<<m$, $(m_{1}\approx m_{2}\approx m)$, and after some calculation, one can find the evolution of stokes parameters similar to (\ref{DDM26}-\ref{DDM28}) except that $\dot{\tau}_{d}$ is defined as follows
	\begin{equation}\label{majorana18}
	\dot{\tau}_{d}=\,(\frac{k^{0}}{\delta})^{2}\frac{3}{8\pi}\,\,(\frac{m_{e}}{m_{d}})^2 \,\,\frac{\sigma_{T}}{\alpha^2} \sqrt{2\pi\langle\sigma v\rangle}\,\, \rho_{d} ,
	\end{equation}
	and so
	\begin{equation}\label{majorana19}
	\frac{\dot{\tau}_{d}}{\dot{\tau}_{e}}=\,\frac{3}{8\pi}\,\,(\frac{k^{0}}{\delta})^{2}(\frac{m_{e}}{m_{d}})^2 \,\,\frac{\Omega_{d}}{\Omega_{B.M}} \,\, \frac{m_{p}}{\alpha^2}\,\, \sqrt{2\pi\langle\sigma v\rangle}.
	\end{equation}
	The above relation clearly shows that the contribution of photon-dark matter scattering on the CMB polarization will be negligble and therefore in the following we will just study the case $\delta<<k^{0}$.
	
\section{GENERALIZED BOLTZMANN EQUATION FOR THE CMB}

The CMB polarization pattern includes two types of polarization, E and B-modes. While the E-mode polarization can be produced via scalar perturbations, the B-mode polarization is only generated by tensor perturbations. In the previous section we showed that the photon-DDM interaction can act as a source for generating B mode in the presence of scalar perturbations. The CMB radiation transfer is described by the multipole moments of temprature $(I)$ and polarization $(P)$ \cite{stoke4,stoke5}
 \begin{equation}\label{BE1}
 \Delta^{S}_{I,P}(\eta,K,\mu)=\sum_{l=0}^{\infty}(2l+1)(-i)^l\Delta^{S}_{I,P_{l}}(\eta,K)P_{l} (\mu),
 \end{equation}
 where $P_{l}(\mu)$ is the Legendre polynomial of rank $l$, $\mu=\hat{\bf{n}}.\hat{\bf{K}}= \cos\theta$ and $\theta$ is the angle between the CMB photon direction $\hat{\bf{n}}=\frac{\bf{k}}{|\bf{k}|}$  and the wave vectors $\bf{K}$ of the Fourier modes of scalar perturbations (S). Since for a given Fourier mode, one can choose a coordinate system in which ${\bf{K}\parallel\hat{\bf{z}}}$ then the Boltzmann equation in the presence of Thomson scattering and DDM-photon interaction can be written as
 \begin{equation}\label{BE2}
 \dfrac{d}{d\eta}\Delta^{S}_{I}+ik\mu\Delta^{S}_{I} +4[\dot{\varPsi}-ik\mu\Phi]=\dot{\tau_{e}}\left[ -\Delta^{S}_{I}+\Delta^{S_{\circ}}_{I}+i\mu v_{b}+\dfrac{1}{2}P_{2}(\mu)\Pi\right],
 \end{equation}
 \begin{equation}\label{BE3}
 \dfrac{d}{d\eta}\Delta^{\pm S}_{P}+ik\mu\Delta^{\pm S}_{P} =\dot{\tau_{e}}\left[ -\Delta^{\pm S}_{P}-\dfrac{1}{2}[1-P_{2}(\mu)]\Pi \right]\mp i a(\eta)\dot{\tau}_{d}\Delta^{\pm S}_{P},
  \end{equation}
  where $\Psi$ and $\Phi$ are the metric perturbations, $\eta$ is the conformal time, $a(\eta)$ is the expansion factor which is normalized to unity for present time $(\eta=\eta_{0})$ and $v_{b}$ is the baryon bulk velocity,  $\Pi\equiv\Delta^{S_{2}}_{I}+\Delta^{S_{2}}_{P}-\Delta^{S_{\circ}}_{P}$ and the polarization anisotropy is given by
  \begin{equation}\label{BE4}
  	\Delta^{\pm S}_{P}=Q^{S}\pm i U^{S},
  \end{equation}
  which can cast the equation of polarization anisotropy into \cite{neutrino}
  \begin{equation}\label{BE5}
 \dfrac{d}{d\eta}\left[ \Delta^{\pm S}_{P}e^{iK\mu\eta\pm i\tilde{\tau}_{d}+\tilde{\tau}_{e}} \right]=-\dfrac{1}{2}e^{iK\mu\eta\pm i\tilde{\tau}_{d}+\tilde{\tau}_{e}}\dot{\tau_{e}}[1-P_{2}(\mu)]\Pi,
  \end{equation}
  where
  \begin{equation}\label{BE6}
  	\tilde{\tau}_{d}(\eta,\mu)\equiv\int_{0}^{\eta}d \eta a(\eta)\dot{\tau}_{d},\,\,\,\,\,\,\,\,\,\,\,\,\tilde{\tau}_{e}(\eta)\equiv\int_{0}^{\eta}d\eta\dot{\tau}_{e}.
  \end{equation}
 Now by integrating (\ref{BE5}) along the line of sight up to the present time $\eta_{0}$, with the initial condition $	\Delta^{\pm S}_{P}(0,K,\mu)=0$, yields
  \begin{equation}\label{BE7}
  \Delta^{\pm S}_{P}(\eta_{0},K,\mu)=\dfrac{3}{4}(1-\mu^{2})\int_{\circ}^{\eta_{\circ}} d\eta\, e^{ix\mu\pm i\tau_{d}(\eta)-\tau_{e}}\,\dot{\tau_{e}}\,\Pi(\eta,K),
  \end{equation}
  where $x=K(\eta_{0}-\eta)$ and
  \begin{equation}\label{BE8}
  	\tau_{d}(\eta)=\int_{\eta}^{\eta_{0}}d\eta a(\eta) \dot{\tau}_{d}(\eta)= \int_{\eta}^{\eta_{0}}d\eta a(\eta)\,\, \sqrt{2\pi\langle\sigma v\rangle}\,\,\frac{\rho _{d}}{m_{d}^2} ,
  \end{equation}
 or in terms of the redshift $z$
  \begin{equation}\label{BE9}
  \tau_{d}(z)=\,\frac{\sqrt{2\pi\langle\sigma v\rangle}}{m_{d}^2}\int_{0}^{z}dz^{'}\,\rho _{d}^{0}\dfrac{(1+z^{'})^{2}}{H(z^{'})},
  \end{equation}
 To obtain (\ref{BE9}) from (\ref{BE8}), we have used the mass density of dark matter $\rho _{d}=\rho _{d}^0 (1+z)^3$ where $\rho _{d}^0$ is mass density of dark matter in present time,  $ad\eta=-\frac{dz}{H(1+z)}$ and by using Friedmann equation in the matter dominated era it can be found as
  \begin{equation}\label{BE10}
  \frac{H^2}{H_{0}^2}=\Omega_{M}^0(1+z)^3+\Omega_{\Lambda}^0,
  \end{equation}
  where $H_{0}\approx67 Kms^{-1}Mpc^{-1}$,\,\,$\Omega_{M}^0\approx0.31$, \,\,$\Omega_{\Lambda}^0\approx0.69$ \cite{early un}.

\begin{figure}
    \centering
    \begin{subfigure}[b]{0.4\textwidth}
        \includegraphics[width=\textwidth]{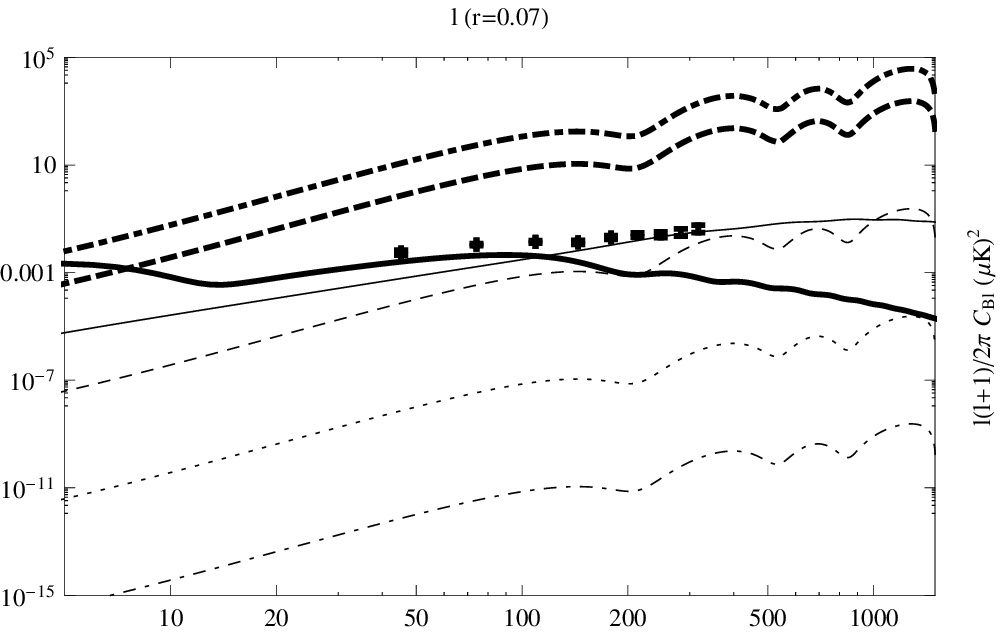}
        \caption{\tiny{To plot above curves, $<\sigma\,v>\simeq10^{-30}cm^3 s^{-1}$ and $r=0.07$ are considered.}}
        \label{DDM287}
    \end{subfigure}
    ~ 
    \begin{subfigure}[b]{0.4\textwidth}
        \includegraphics[width=\textwidth]{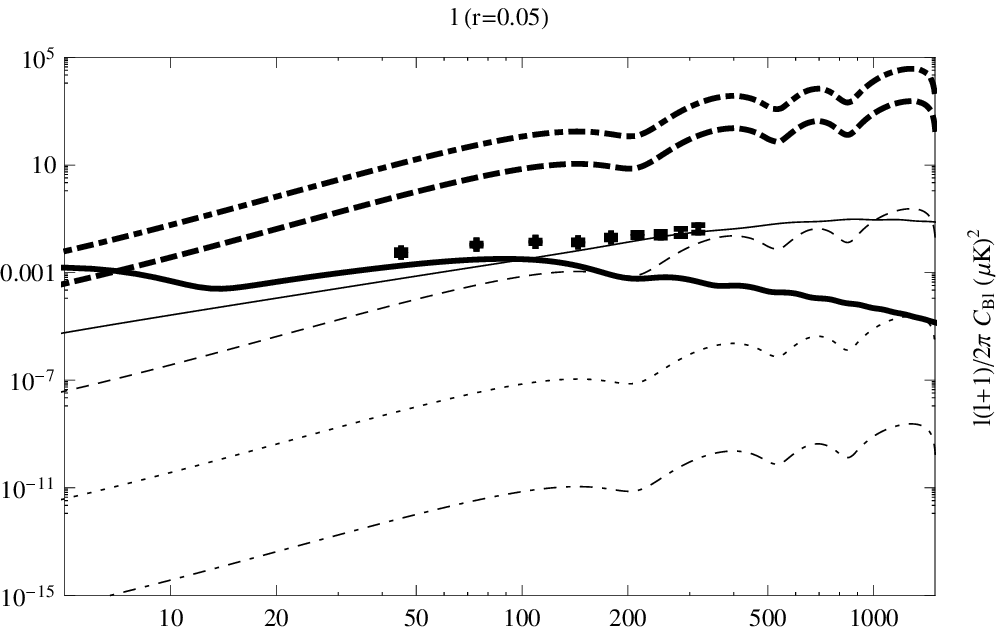}
        \caption{\tiny{To plot above curves, $<\sigma\,v>\simeq10^{-30}cm^3 s^{-1}$ and $r=0.05$ are considered.}}
        \label{DDM285}
    \end{subfigure}
    \par

    ~ 
    \caption{\tiny{The magnetic like linear polarization angular power spectrum $l (l+1)/(2\pi)\,C^{BB}_l$ for different values of the tensor to scalar ratio $r$ is plotted in terms of $(\mu K)^2$;  narrow, thick, thick-dashed-dotted, thick-dashed, dashed, dotted and dashed-dotted lines indicate $C^{BB}_l$ due to: the gravitation lensing effects, the standard contribution due to Compton scattering in the presence of tensor perturbations with $r$ mentioned in sub-caption,  the Dark matter magnetic moment contribution in the presence of scalar metric perturbations with different masses $m_D\equiv\{1MeV,5MeV,100MeV,500MeV,1GeV\}$, respectively. The points with error bars show the BICEP2/Keck Array data. We have chosen the Planck best fit values for the cosmological parameters.}}\label{ClB-1}
     \begin{subfigure}[b]{0.4\textwidth}
     	\includegraphics[width=\textwidth]{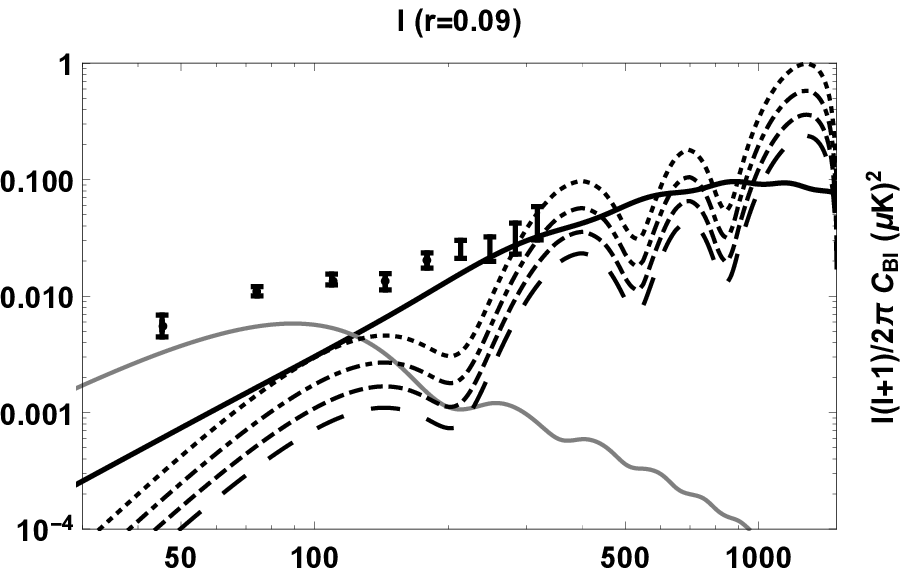}
     	\caption{\tiny{To plot above curves, $<\sigma\,v>\simeq10^{-30}cm^3 s^{-1}$ and $r=0.09$ are considered.}}
     	\label{DDM286}
     \end{subfigure}
     \caption{\tiny{The magnetic like linear polarization angular power spectrum $l (l+1)/(2\pi)\,C^{BB}_l$ for is plotted in terms of $(\mu K)^2$;  thick black, thick gray ,dotted, dashed-dotted, dashed, big dashed lines indicate $C^{BB}_l$ due to: the gravitation lensing effects, the standard contribution due to Compton scattering in the presence of tensor perturbations,  the Dark matter magnetic moment contribution in the presence of scalar metric perturbations with different masses $m_D\equiv\{70MeV,80MeV,90MeV,100MeV\}$, respectively. The points with error bars show the BICEP2/Keck Array data. We have chosen the Planck best fit values for the cosmological parameters.}}\label{ClB-2}
\end{figure}

  Meanwhile E-mode and B-mode polarizations can be defined in terms of $\Delta^{\pm S}_{P}(\eta_{0},K,\mu)$ as follows \cite{stoke1,stoke3,stoke4,stoke5}
  \begin{equation}\label{BE11}
  \Delta^{(S)}_{E}(\eta_{0},K,\mu)\equiv-\dfrac{1}{2}\left[\bar{\eth}^{2}\,\Delta^{+(S)}_{P}(\eta_{0},K,\mu)+\,\eth^{2}\Delta^{-(S)}_{P}(\eta_{0},K,\mu) \right],
  \end{equation}
  \begin{equation}\label{BE12}
  \Delta^{S}_{B}(\eta_{0},K,\mu)\equiv\dfrac{i}{2}\left[\bar{\eth}^{2}\,\Delta^{+S}_{P}(\eta_{0},K,\mu)-\eth^{2}\Delta^{-S}_{P}(\eta_{\circ},K,\mu) \right],
  \end{equation}
  where $\eth $ and $\bar{\eth}$ are spin raising and lowering operators, respectively. Thus by assumming the scalar perturbation to be axially symmetric around $\bf{K}$ one has
  \begin{equation}\label{BE13}
  \bar{\eth}^{2}\Delta^{+S}_{P}(\eta_{0},K,\mu)=\partial^{2}_{\mu}\left[(1-\mu^{2}) \Delta^{+S}_{P}(\eta_{0},K,\mu) \right],
  \end{equation}
  \begin{equation}\label{BE13}
  {\eth}^{2}\Delta^{-S}_{P}(\eta_{0},K,\mu)=\partial^{2}_{\mu}\left[(1-\mu^{2}) \Delta^{-S}_{P}(\eta_{0},K,\mu) \right],
  \end{equation}
which can cast (\ref{BE11}) and (\ref{BE12}) into
  \begin{equation}\label{BE14}
  \Delta^{S}_{E}(\eta_{0},K,\mu)=-\dfrac{3}{4}\int_{0}^{\eta_{0}} d\eta\,g_{e}(\eta)\Pi(\eta,K)\partial^{2}_{\mu}[(1-\mu^{2})e^{ix\mu}\cos\tau_{d}],
  \end{equation}
  \begin{equation}\label{BE15}
  \Delta^{S}_{B}(\eta_{0},K,\mu)=\dfrac{3}{4}\int_{0}^{\eta_{0}} d\eta\,g_{e}(\eta)\Pi(\eta,K)\partial^{2}_{\mu}[(1-\mu^{2})e^{ix\mu}\sin\tau_{d}],
  \end{equation}
  where $g_{e}(\eta)=\dot{\tau}_{e} e^{-\tau_{e}}$ is the visibility function of electron. As one can easily see for $\tau_{d}\neq0$ the equations (\ref{BE14}) and (\ref{BE15}) show that  the DDM-photon interaction produces the nontrivial B-mode polarization and modify of the ordinary E-mode polarization.
  However, the power spectrum for the E and B-modes can be obtained by integrating over the initial power spectrum of the metric perturbation as \cite{stoke1,stoke3,stoke4,stoke5}
  \begin{equation}
  C^{EE,S}_{l}=\frac{1}{2l+1}\dfrac{(l-2)!}{(l+2)!}\int d^{3}K\,P_{S}(K) \left|  \sum_{m}\int d\Omega\,Y_{lm}^{*}({\bf{n}})\Delta^{S}_{E}(\eta_{0},K,\mu) \right| ^{2},\label{Emode}
  \end{equation}
  \begin{equation}
  C^{BB,S}_{l}=\frac{1}{2l+1}\dfrac{(l-2)!}{(l+2)!}\int d^{3}K\,P_{S}(K) \left|  \sum_{m}\int d\Omega\,Y_{lm}^{*}({\bf{n}})\Delta^{S}_{B}(\eta_{0},K,\mu) \right| ^{2},\label{Bmode}
  \end{equation}
  where $P_{S}(K)$ is the initial power spectrum of the scalar mode perturbation. The equations (\ref{Emode}) and (\ref{Bmode}) can be rewritten as
  \begin{equation}\label{CLE}
  C^{EE,S}_{l}=(4\pi)^{2}\dfrac{(l+2)!}{(l-2)!}\int d^{3}K\,P_{S}(K) \left|\dfrac{3}{4}  \int_{0}^{\eta_{0}} d\eta\, g_{e}(\eta)\,\Pi(\eta,K)\dfrac{j_{l}}{x^{2}}\,\cos\tau_{d} \right| ^{2},
  \end{equation}
  \begin{equation}\label{CLB}
  C^{BB,S}_{l}=(4\pi)^{2}\dfrac{(l+2)!}{(l-2)!}\int d^{3}K\,P_{S}(K) \left|  \dfrac{3}{4}\int_{0}^{\eta_{0}} d\eta\,g_{e}(\eta)\,\Pi(\eta,K)\,\dfrac{j_{l}}{x^{2}}\,\sin\tau_{d}\right| ^{2},
  \end{equation}
  by using identities $\partial_{\mu}^2 (1-\mu^2)e^{ix\mu}\equiv(1+\partial_{x}^2)x^2 e^{ix\mu}$ and $\int d\Omega Y_{lm}^* e^{ix\mu}=\\(i)^l\sqrt{4\pi(2l+1)j_{l}(x)}\delta_{m0} $.
We have numerically calculated the B-mode power spectra using CMBquick code for different values of $\left<\sigma v\right>$, $m_d$ and $r$-parameter as are shown in Figs.(\ref{ClB-1}) and (\ref{ClB-2}). However to see how the curves depends on the dark matter one can approximate (\ref{CLE}) and (\ref{CLB}) as follows
  \begin{equation}
  C^{EE,S}_{l}= \bar{C}^{EE,S}_{l}(\cos^{2}\bar{\tau}_{d}),
  \end{equation}
  and
  \begin{equation}
  C^{BB,S}_{l}= \bar{C}^{EE,S}_{l}(\sin^{2}\bar{\tau}_{d})\label{Bmodeav},
  \end{equation}
  where $\bar{C}^{EE,S}_{l}$ is the value of the power spectrum for E mode polarization associated with the Compton scattering in the context of scalar perturbation \cite{stoke4,stoke5}
  \begin{equation}
 \bar{C}^{EE,S}_{l}=(4\pi)^{2}\dfrac{(l+2)!}{(l-2)!}\int d^{3}K\,P_{S}(K) \left|\dfrac{3}{4}  \int_{0}^{\eta_{0}} d\eta\, g_{e}(\eta)\,\Pi(\eta,K)\dfrac{j_{l}}{x^{2}}\right| ^{2},
  \end{equation}
  and $\bar{\tau}_{d}$ is the time average of $\tau_{d}$

  \begin{eqnarray}\label{taubar}
 \bar{\tau}_{d}(z)&=&\frac{1}{z_{l}}\int_{0}^{z_{l}}dz\,\, \tau_{d}(z)\,\,\approx
 \,\,\frac{1.4 \times 10^{-1}
MeV	^2 }{m_{d}^2}\,\,\,\,\,\,\nonumber\\
&&(\frac{\langle\sigma v\rangle}{10^{-30}\frac{cm^3}{s}}) \,\,\,(\frac{\rho _{d}^0}{2.5\times10^{-30} \frac{gr}{cm^{3}}})\,\,,
  \end{eqnarray}

  with $z_{l}\approx1100$ represents redshift at the last scattering surface. Therefore, one can expect a valuable contribution on the $ C^{BB,S}_{l}$ from dark matter with mass about a few 10MeV.

\begin{figure}
      \centering
    \begin{subfigure}[b]{0.4\textwidth}
        \includegraphics[width=\textwidth]{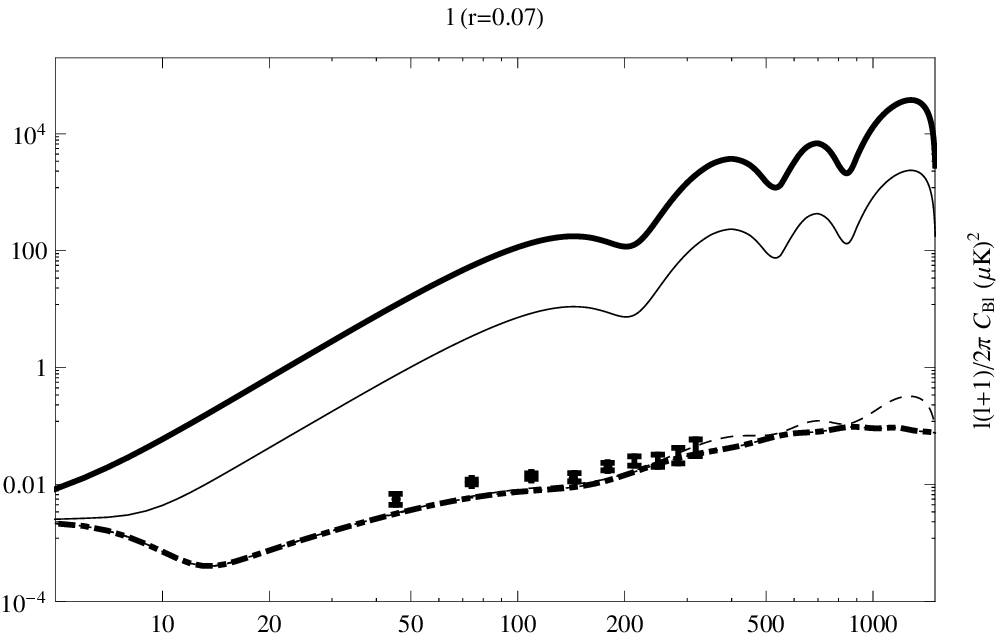}
        \caption{\tiny{To plot above curves, $<\sigma\,v>\simeq10^{-30}cm^3 s^{-1}$ and $r=0.07$ are considered.}}
        \label{T287}
    \end{subfigure}
    ~ 
    \begin{subfigure}[b]{0.4\textwidth}
        \includegraphics[width=\textwidth]{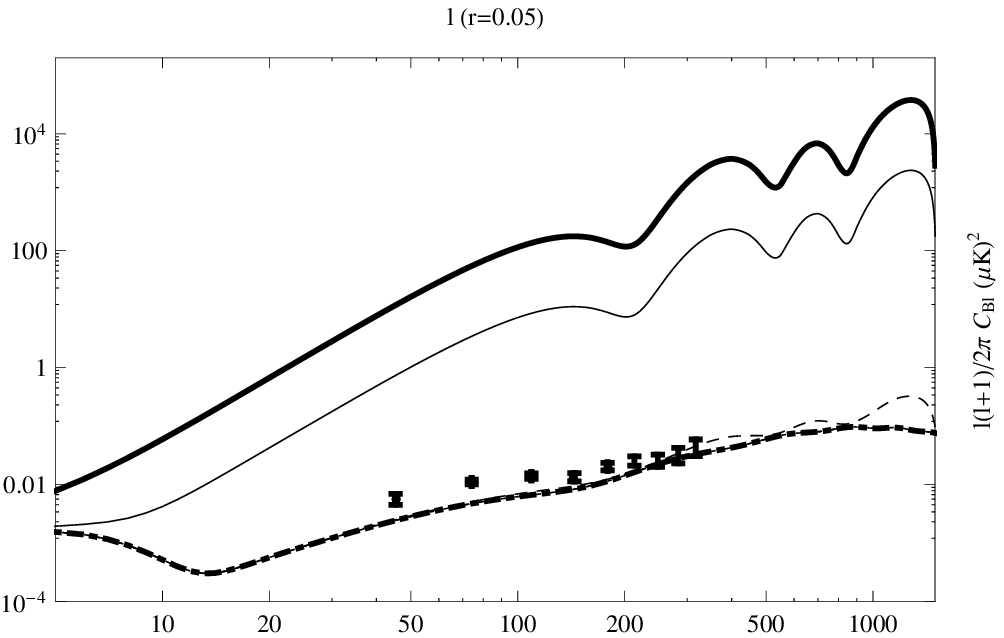}
        \caption{\tiny{To plot above curves, $<\sigma\,v>\simeq10^{-30}cm^3 s^{-1}$ and $r=0.05$ are considered.}}
        \label{T288}
    \end{subfigure}
    \par

    ~ 
    \caption{\tiny{The magnetic like linear polarization angular power spectrum $l (l+1)/(2\pi)\,C^{BB}_l$ for different values of the tensor to scalar ratio $r$ and $<\sigma\,v>$ is plotted in terms of $(\mu K)^2$; the diagrams show: Compton scattering in presence of the tensor perturbations and gravitation lensing effect without considering DDM interactions (thick-dashed line), Compton scattering in the presence of the tensor perturbations and gravitation lensing plus DDM interactions in the presence of scalar perturbations  by considering different masses for dark matter $m_D=1MeV$ (thick line), $m_D=5MeV$ (narrow line), $m_D=100MeV$ (dashed line), $m_D=500MeV$ (dotted line) and $m_D=1GeV$ (dashed-dotted line) respectively. The points with error bars show the BICEP2/Keck Array data. We have chosen the Planck best fit values for the cosmological parameters.}}\label{ClB-3}
     \begin{subfigure}[b]{0.4\textwidth}
     	\includegraphics[width=\textwidth]{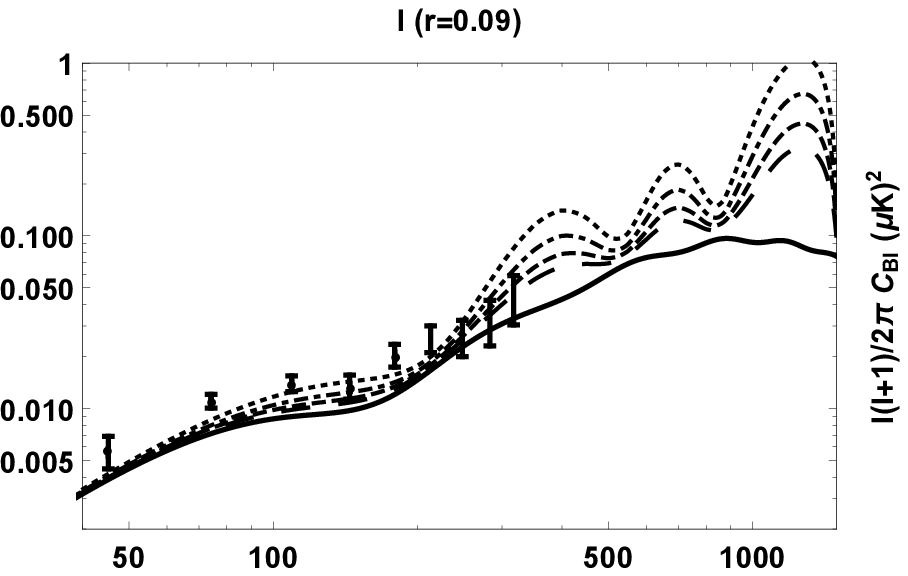}
     	\caption{\tiny{To plot above curves, $<\sigma\,v>\simeq10^{-30}cm^3 s^{-1}$ and $r=0.09$ are considered.}}
     	\label{T289}
     \end{subfigure}
      \caption{\tiny{The magnetic like linear polarization angular power spectrum $l (l+1)/(2\pi)\,C^{BB}_l$ is plotted in terms of $(\mu K)^2$; The plot shows: Compton scattering in presence of the tensor perturbations with $r=0.09$ and gravitation lensing effect without considering DDM interactions (thick line), Compton scattering in the presence of the tensor perturbations and gravitation lensing plus DDM interactions in the presence of scalar perturbations  by considering different masses for dark matter $m_D=70MeV$ (dotted line), $m_D=80MeV$ (dashed-dotted line), $m_D=90MeV$ (dashed line), $m_D=100MeV$ (big-dashed line), respectively. The points with error bars show the BICEP2/Keck Array data. We have chosen the Planck best fit values for the cosmological parameters.}}\label{ClB-4}
\end{figure}

\section{ Conclusion and discussion}
Producing of the magnetic-like linear polarization power spectrum of the CMB photons is estimated by using  Quantum
Boltzmann Equation. For this purpose the CMB radiation is considered as an ensemble of photons which is described by the Stokes parameters.
Meanwhile, the Compton scattering and photon-dark matter dipolar interaction are considered as the collision terms in Quantum
Boltzmann Equation.
As shown in Eqs.(\ref{Bmode}) and (\ref{Bmodeav}), the power spectrum of the
 B-mode of CMB polarizations are modified in the presence of CMB-DDM interaction.  The most important point is that the B-mode is generated by the
CMB-DDM interaction in the presence of the scalar perturbation which is in contrast with the standard scenario for the generation of  the CMB B-mode.
It should be emphasized that the
$r$ ratio is usually introduced by
comparing B- and E-modes linear polarization power spectrum while it is assumed that the observed B-mode $C^{ob}_{Bl}$ is totally attributed to
Thomson scattering in the presence of tensor perturbations  $C^{ob}_{Bl} =C^T_{Bl}$.
However, our results show that other interactions such as CMB-DDM interaction  can generate
magnetic like power spectrum in the presence of scalar perturbations $C_{Bl}^{(S)}$ and modify the $r$-parameter as follows
\begin{equation}\label{r01}
    \rm r\propto C^T_{Bl}/C^{(S)}_{El}\propto (C^{ob}_{Bl}-C^{(S)}_{Bl})/C^{(S)}_{El}.
\end{equation}
As (\ref{r01}) shows, the value of the $r$-parameter, as a scale of the amplitude of gravitational wave, is suppressed. By using (\ref{Emode}-\ref{taubar}) in  (\ref{r01}), one can approximately find
 \begin{equation}\label{r02}
    r\simeq r^{0}-\sin^2\bar\tau_d,
\end{equation}
where $r^0$ is the standard tensor to scalar perturbation ratio without considering any new source for the B-mode polarization such as the CMB-DDM interaction. \par
However, the value of  B-mode power spectrum in the presence of scalar perturbation and CMB-DDM interaction $C_{Bl}^{(S)}$ depends on
$\tau_d$ (\ref{BE8}) as is shown in (\ref{CLB}) and (\ref{Bmodeav}) . To compare the contribution of CMB-DDM interaction with respect to the Compton scattering, we need to have the ratio $\frac{\dot\tau_d}{\dot\tau_e}$ or equally $(\frac{m_e}{m_d})^2$ and $\sqrt{<\sigma\,v>}$. To this end
the numerical value of the B-mode power spectrum of the CMB for different values of $m_{d}$ and $r$ for $<\sigma\,v>\approx10^{-30}cm^3 s^{-1}$  together with BICEP2/Keck Array data is plotted in Figs. (\ref{ClB-1}), (\ref{ClB-2}), (\ref{ClB-3}) and (\ref{ClB-4}). Figs. (\ref{ClB-1}) and (\ref{ClB-2}) show the behaviour of the B-mode power spectrum due to Compton scattering in the presence of tensor perturbations, the gravitational lensing effects and photon-DDM interaction in the case of scalar perturbation. On the other hand, Figs (\ref{ClB-3}) and (\ref{ClB-4}) indicate the contribution of lense effects and compton scattering in the presence of tensor perturbation plus photon-DDM interaction. As the figures show for different values of the $r$ parameter one can obtain different mass regions for DDM. For example, the contribution of B-mode due to CMB-DDM interaction in the presence of scalar perturbations for $r=0.07$ and $m_d\leq 50MeV$ ($r=0.09$ and $m_d\leq80MeV$ ) is larger than total reported B-mode power spectrum and therefore can be excluded experimentally. In fact the B-mode polarization power spectrum can put a bound on the magnetic dipole moment about $M\leq 10^{-16} e\,\,cm$  which is in agreement with other reported constraints \cite{DDM.EM,constraint1,constraint2,constraint3}.

To summarize, it should be noted that, in this paper, we have studied the effects of Dirac DDM-photon interaction as well as the effects of DDM-photon interaction which occur through Majorana magnetic moment in the case of $\delta=m_{2}-m_{1}\ll k^{0}$. If $\delta\gg k^{0}$, one can find that the similar results  to the previous one ($\delta<<k^{0}$) will be obtained except that the solution will be suppressed as $(\frac{k^{0}}{\delta})^{2}$. Also, it is important to note that in the case of $\delta=k^{0}$, the resonance mode will be occurred which is under investigation as a future work \cite{future}. As a final point, we should mention that the future observed data for B-mode polarization power spectrum can be used as an indirect probe of the nature of photon-dark matter interaction. \par


\end{document}